\documentclass{article}[11]

\usepackage{graphics,latexsym}
\usepackage{graphicx}
\usepackage{amsmath, amssymb,amsthm}

\begin{document}

\title{Trapping Electromagnetic Solitons in Cylinders}

\author{Daniele Funaro}

\date{}
\maketitle

\centerline{Dipartimento di Fisica, Informatica e Matematica}
\centerline{Universit\`a di Modena e Reggio Emilia}
\centerline{Via Campi 213/B, 41125
Modena (Italy)} \centerline{daniele.funaro@unimore.it}

\begin{abstract} {\small Electromagnetic waves, in vacuum or dielectrics,
can be confined in unbounded cylinders in such a way that they turn around 
the main axis. For particular choices of the cylinder's section, interesting
stationary configurations may be assumed. By refining some results obtained 
in previous papers, additional more complex situations are examined here. 
For such peculiar guided waves an explicit expression is given in terms of 
Bessel's functions. Possible applications are in the development of whispering 
gallery resonators.}
\end{abstract}

\vspace{.1cm}
\par\smallskip


\section{Introduction}

The classical equations of electromagnetism
allow for solutions confined in ring-shaped domains. In vacuum, this is
made possible by the orthogonality of the electric and magnetic
fields (${\bf E}$ and ${\bf B}$) and by the enforcement of the
two divergence-free conditions (${\rm div}{\bf E}=0$ and
${\rm div}{\bf B}=0$). In fact, if the lines of force are
closed and orthogonal, a toroid is a natural environment to
set up the initial conditions, that successively evolve
according to:
\begin{equation}\label{max}
\frac{\partial {\bf E}}{\partial t}~=~ c^2{\rm curl} {\bf B}
\hspace{1.5cm}
\frac{\partial {\bf B}}{\partial t}~=~ -{\rm curl} {\bf E}
\end{equation}
where $c$ denotes the speed of light.
The above time-dependent equations are easily put in relation with
the vector wave equation and its corresponding eigenmodes.
\smallskip

The search for electromagnetic waves trapped in a toroid poses
interesting mathematical questions. Numerical computations show a
variety of solutions, whose dynamics depends on the
section's shape. The behavior is strikingly similar to that of a
non-viscous fluid confined in a vortex ring, but with additional
intrinsic constraints.
\smallskip

Explicit full solutions in terms of Bessel functions are available
in the case of cylinders,
where the magnetic field oscillates parallel to the axis and the
electric field lays on the circular sections. The configuration
recalls that of a train of solitons smoothly circulating inside a
2D rounded cavity. As documented in \cite{chinosi}, thin rings with
large diameter and circular section can be
well approximated by the above mentioned solutions.
\smallskip

For more compact rings the use of numerical simulations is
a necessity. By the way, not all the shapes are workable. Indeed,
only a restricted range of sections are compatible with the
electromagnetic constraints. Thus, the solution process must be
implemented together with a sort of shape-detection algorithm
(see \cite{chinosi}).
\smallskip

Extensions of previous results (briefly recalled in the next section)
are here obtained for electromagnetic
waves trapped in unbounded cylinders where, the rotation around the
axis is combined with a radial oscillation. Here we assume that the
section is an annulus, so that the corresponding domain is a hollow cylinder.
Having in mind the vector wave
equation, the study is connected to the search of eigenfunctions
of the Laplace operator in such a way  that the dimension of the corresponding
eigenspace is equal to four. As we will see,
this analysis leads to the study of specific properties of Bessel's
functions.

\section{Electromagnetic waves turning around an axis}

We assume that $k\geq 1$ is an integer number.
We recall that the $k$-th Bessel's functions, of the first and the second
kinds respectively, are independent solutions to the same eigenvalue
problem ($r>0$):
\begin{equation}\label{besselj}
J_k^{\prime\prime}(\omega r)~+~\frac {J_k^\prime (\omega r)}
{ r}~-~\frac{k^2J_k(\omega r)} {r^2}~= ~-\omega^2 J_k(\omega r )
\end{equation}
\begin{equation}\label{bessely}
Y_k^{\prime\prime}(\omega r)~+~\frac {Y_k^\prime (\omega r)}
{ r}~-~\frac{k^2Y_k(\omega r)} {r^2}~=~ - \omega^2 Y_k(\omega r )
\end{equation}
for a given parameter $\omega$.
We also recall that $J_k$ tends to zero for $r\rightarrow 0$, 
while $Y_k$ is unbounded in the neighborhood of $r=0$
(see, e.g., \cite{watson}).
\smallskip

Solutions of the entire set of Maxwell's equations are obtained
in cylindrical coordinates $(r,z, \phi )$ as follows (see \cite{funaro},
\cite{funaro2}):
$${\bf E}=\left(
\frac {k f(\omega r)} {\omega r}\cos (c\omega t-k\phi)
,~~0,~ f^\prime (\omega r)\sin (c\omega
t-k\phi)\right)$$
\begin{equation}\label{cbdiskk}
{\bf B}~=~\frac {1} {c}\Big( 0,~
f(\omega r) \cos (c\omega t-k\phi),~~0\Big)
\end{equation}
where $f$ is a linear combination of $J_k$ and $Y_k$.
The magnetic field is parallel to the cylinder's axis and
the electric field belongs to the plane $(r, \phi )$.
Bounded solutions are possible in two cases: the domain
is such that $r\geq R_m >0$ for some given $R_m$ (see for instance
the second picture in Fig. 1); the function $f$ is just a multiple of
$J_k$, thus $r=0$ may be included in the domain (see for instance
the first picture in Fig. 1).
\smallskip

The displacement of the electric field at a given time is
shown in Fig. 1 for $k=1$. During time evolution, the
electromagnetic wave rigidly rotates around the central point.
Two examples are taken into account. In the left picture,
the domain is the circle $0\leq r\leq R_M$, where
$R_M$ is the first zero of $J_1$, which takes approximately
the value 3.832. In this way, at the boundary, both the
magnetic field and the radial component of the electric
field are zero. In the right picture, we have the annular
region $R_m=1\leq r\leq R_M$. The outer diameter is now chosen
in such a way that the electric field is orthogonal to the
boundary (perfect conductivity).

\begin{center}
\vspace{-.1cm}
\begin{figure}[h]
\centerline{\includegraphics[width=8.4cm,height=6.5cm]{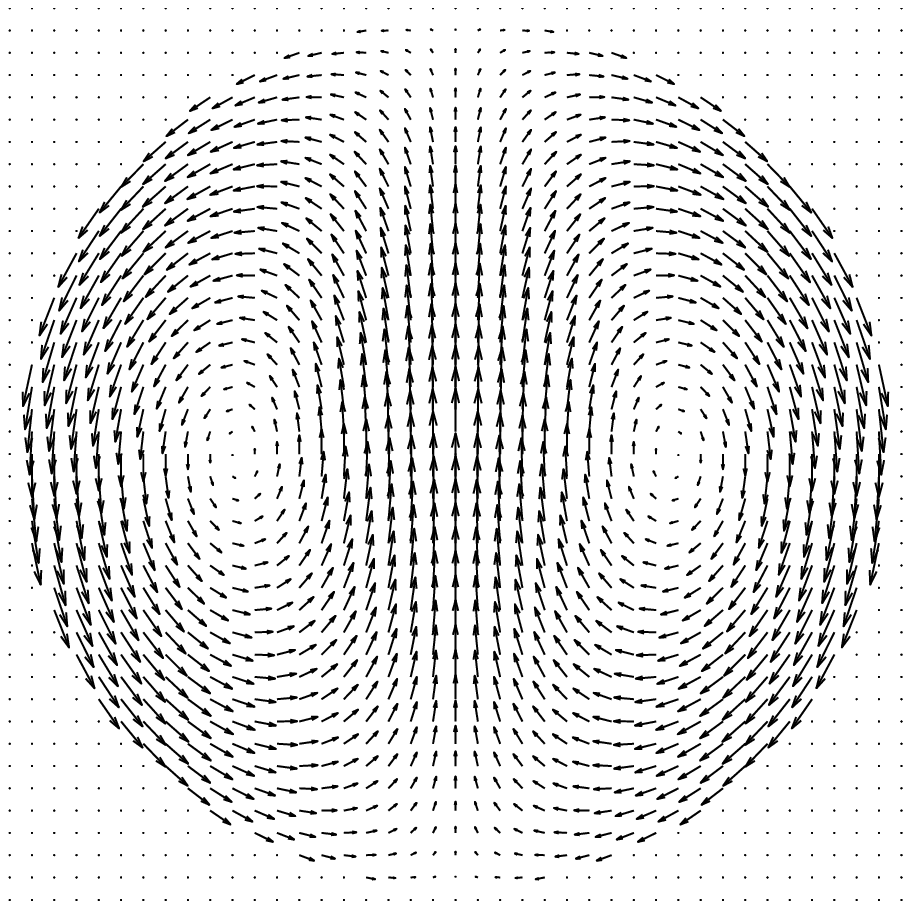}
\hspace{-1.7cm}\includegraphics[width=8.4cm,height=6.5cm]{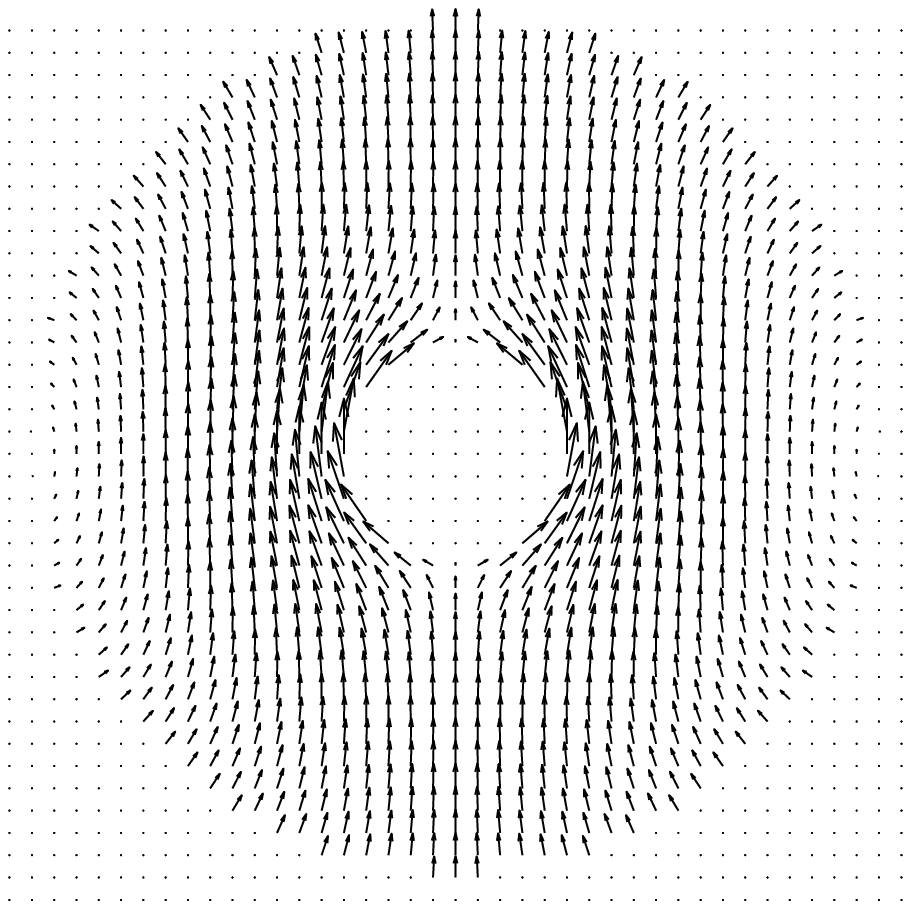}}
\vspace{-.2cm}
\begin{caption}{\sl Displacement of the electric field at fixed time
obtained from the expression in (\ref{cbdiskk}) subject to different kinds
of boundary constraints. The sizes of the domains are not arbitrary,
but strictly related to the frequency of rotation as well as boundary conditions.
The magnetic field is orthogonal to the page.}
\end{caption}
\end{figure}
\end{center}\vspace{-.1cm}

Once the frequency of rotation (depending on $c\omega$) is fixed
and the type of boundary conditions decided, the size of the domain
is automatically constrained. This means that there are few
domains allowing for the creation of these waves. If the proper
size is not respected, the wave will not follow a stationary
motion, i.e., it does not complete the cycle preserving the
phase. The conditions that permit such a guided behavior depend
on the eigenvalues of the Laplace operator on the domain. If
the shape is such that some eigenvalues have multiplicity higher
than one, then the construction of such rotating solitary waves is
possible (we show later the determination process).
An analysis of spherical vortices has been provided in
\cite{chinosi}, exactly with the purpose of detecting domains
where two independent eigenfunctions are related to the same
eigenvalue. In this paper, we stay for simplicity in the case of the cylinder
by analyzing the peculiar case where four independent eigenfunctions share the same
eigenvalue. The evolution in a thin toroid can be approximated, with a rather good
level of accuracy, starting from the cylinder's version;
this extension however is not going to be studied here.

\section{A more involved evolution}

We would like to analyze the situation in which the multiplicity
of one of the eigenvalues of the Laplace operator, on a certain
annular domain, is equal to four. This will allow us to build
more complex electromagnetic waves turning around an axis.
We will need to play with both the functions $J_k$ and $Y_k$,
therefore we have to stay away from the point $r=0$ (recall
that $Y_k$ is singular there). From now on we assume that
$R_m=1\leq r\leq R_M$, where the major diameter $R_M$ has to be
properly determined. Of course, if $R_m$ is modified the entire
setting scales accordingly due to the linearity of Maxwell's
equations.
\smallskip

We impose Dirichlet boundary conditions on both boundaries (the inner
and the outer circumferences), though other conditions may be considered.
Taking Dirichlet conditions in the inner part involves working
with the function:
\begin{equation}\label{besco}
f_n(r)~=~J_n(\lambda r)Y_n(\lambda )~-~Y_n(\lambda r)J_n(\lambda)
\end{equation}
where $\lambda >0$ is a parameter.
Of course, we have $f_n(1)=0$, $\forall n$ and $\forall \lambda$.
\smallskip

Now, we would like to have:
\begin{equation}\label{sys}
\begin{cases}f_n(R_M)~~=~0 \cr f_m(R_M)~=~0\end{cases}~~~~~~~~~~~~~ m\not =n
\end{equation}
which is a nonlinear problem leading to the detection of $R_M$
and $\lambda$. This means that, once the diameter of the inner
circumference has been fixed, the frequency of the evolving
wave and the magnitude of the entire domain are going to be uniquely determined
by (\ref{sys}). Not necessarily such a problem has solution.
However, with the help of numerical tests, we were able to establish some
facts.
\smallskip

Here below we report some of the conclusions of our analysis:
$$
m=1,~ n=2: ~{\rm no ~ values ~ of}~
R_M~{\rm and}~\lambda~{\rm were~found}
$$
$$
m=1,~ n=3: ~{\rm no ~ values ~ of}~
R_M~{\rm and}~\lambda~{\rm were~found}
$$
$$
m=1,~ n=4: ~~~R_M=7.0927~~~{\rm for}~\lambda =1.0698
%m=1,~ n=4: ~~~R_M=7.09275~~~{\rm for}~\lambda =1.069896
$$
$$
m=1,~ n=5: ~~~R_M=3.6761~~~{\rm for}~\lambda =2.3871
%m=1,~ n=5: ~~~R_M=3.6761278~~~{\rm for}~\lambda =2.387188
$$
$$
m=1,~ n=6: ~~~R_M=2.7603~~~{\rm for}~\lambda =3.6054
%m=1,~ n=6: ~~~R_M=2.7603013~~~{\rm for}~\lambda =3.6054147
$$
$$
m=2,~ n=3: ~{\rm no ~ values ~ of}~
R_M~{\rm and}~\lambda~{\rm were~found}
$$
$$
m=2,~ n=4: ~{\rm no ~ values ~ of}~
R_M~{\rm and}~\lambda~{\rm were~found}
$$
$$
m=2,~ n=5: ~~~R_M=5.1500~~~{\rm for}~\lambda =1.7032
%m=2,~ n=5: ~~~R_M=5.150043~~~{\rm for}~\lambda =1.703224
$$
\newpage

Let us better examine a specific situation ($m=1$, $n=4$).
According to Fig. 2, both functions $f_1$ and $f_4$ are zero
for $r=R_m=1$ and $R_M\approx  7.0927$. Thus, by imposing the same boundary
constraints, we are able to find out two branches of vector solutions of the type
given in (\ref{cbdiskk}) ($k=m$ and $k=n$ respectively)
related to the same frequency of rotation. These are generated by
magnetic fields whose intensity is given by the level lines of Fig. 3.
\smallskip

\begin{center}
\vspace{.2cm}
\begin{figure}[h]
\centerline{\includegraphics[width=9.cm,height=7.cm]{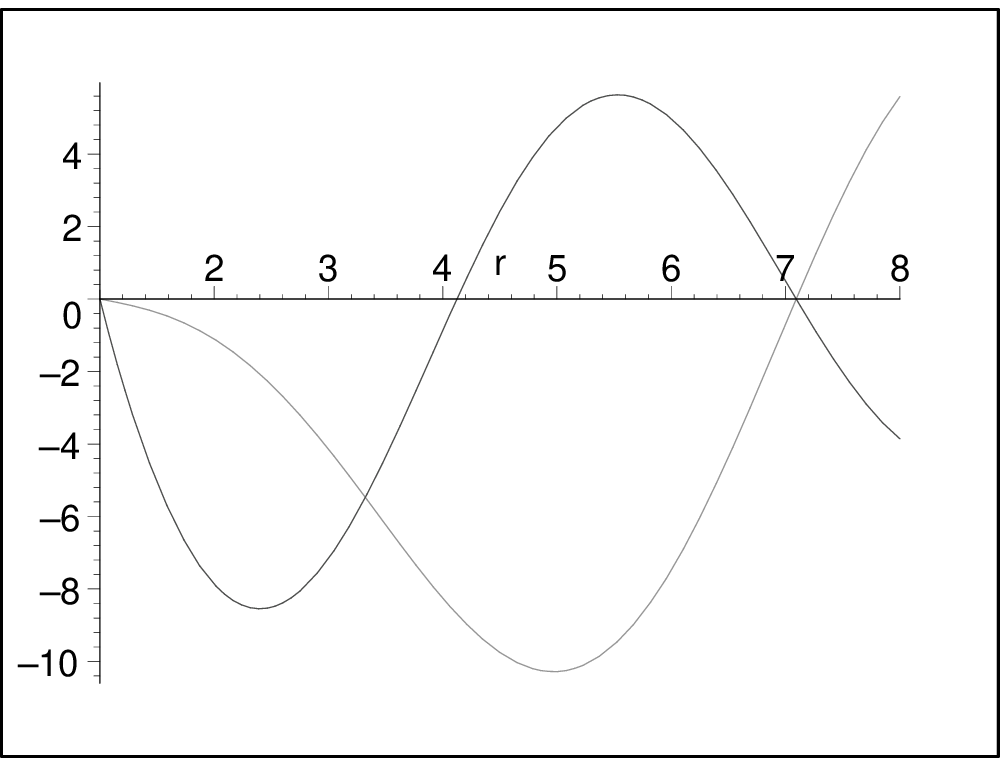}}
\vspace{.1cm}
\begin{caption}{\sl Plots of $f_1$ and $f_4$ for
$\lambda \approx1.0699$. In this case both the functions vanish at $r=1$
and $r\approx 7.0927$.}
\end{caption}
\end{figure}
\end{center}\vspace{.2cm}

Successively, starting from $\lambda$, we
can build general solutions of the form:
\begin{equation}\label{evol}
w_1 \sin(c\sqrt{\lambda} ~t-\phi_0)+w_2 \sin(c\sqrt{\lambda}~ t)+
w_3 \cos(c\sqrt{\lambda}~ t-\phi_0)+w_4 \cos(c\sqrt{\lambda}~ t)
\end{equation}
The function $w_j$, $j=1,2,3,4$, are schematically reported in Fig. 4.
In particular,  $w_1$ and $w_3$ are as shown in the first picture of Fig. 3,
but with a difference of phase of 90 degrees.
Combining $w_1$ and $w_3$ we can obtain the electromagnetic fields
according to the expression (\ref{cbdiskk}) for $k=1$.  
The other two functions, $w_2$ and $w_4$, are as shown in the second
picture of Fig. 3 and differ for an angle of
22.5 degrees. They also lead to (\ref{cbdiskk}) (this time with $k=4$).
The phase lag $\phi_0$ is arbitrarily given.
\smallskip

It is just a direct computation verifying that the expression provided
in (\ref{evol}) solves the wave equation in vacuum.
In fact, the second derivative in time produces the multiplicative
factor $-c^2\lambda$, while the application of the Laplace operator
is equivalent to a multiplication by $-\lambda =-\omega^2$ (see
(\ref{besselj}) and (\ref{bessely})).  
\smallskip

In Fig. 5, one can see, at different time steps, the 
evolution of (\ref{evol}) for $\phi_0=0$. Only half cycle is displayed. From the
last picture the sequence restarts from the beginning with the
color interchanged. An animation is available in \cite{funarow}.
The effect vaguely recalls the juggling of three clubs.

\begin{center}
\vspace{.9cm}
\begin{figure}[h]
\centerline{\includegraphics[width=8.4cm,height=6.5cm]{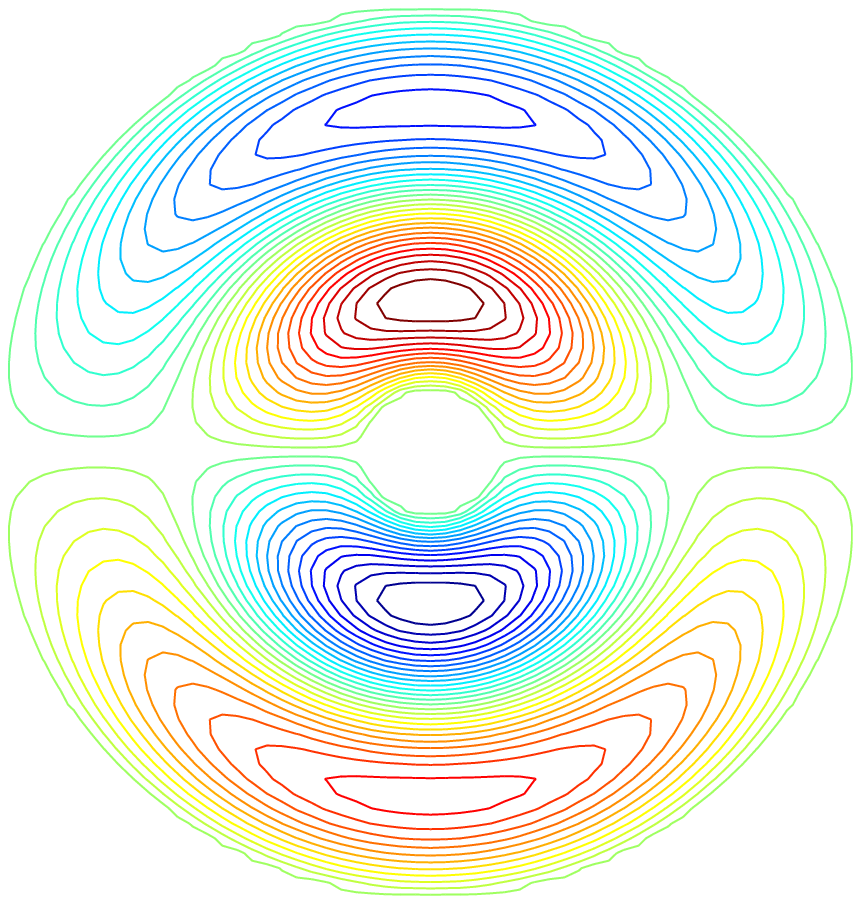}
\hspace{-1.7cm}\includegraphics[width=8.4cm,height=6.5cm]{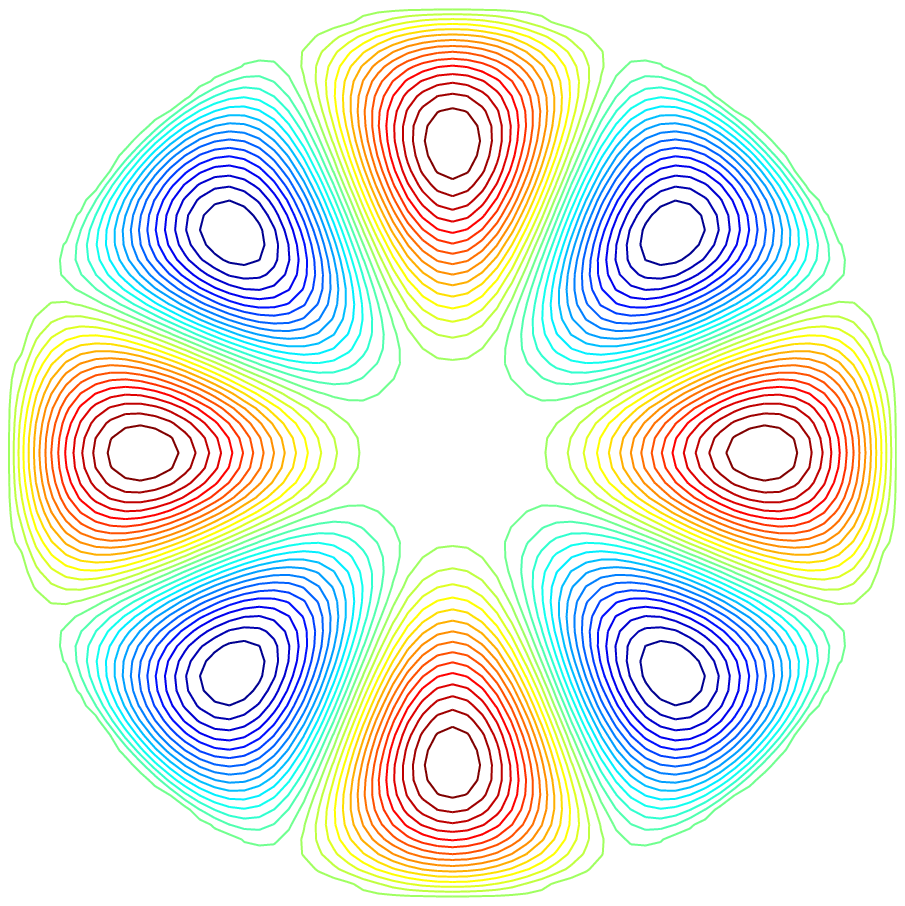}}
\vspace{-.4cm}
\begin{caption}{\sl Plots of two independent eigenfunctions of the
Laplace operator with homogeneous Dirichlet boundary conditions on an
annular domain. Despite the fact that they look totally different, the
domain has been dimensioned in such a way the corresponding eigenvalues are the same.}
\end{caption}
\end{figure}
\end{center}\vspace{.1cm}

\begin{center}
\begin{figure}[h]
\centerline{\hspace{.4cm}
\includegraphics[width=12.cm,height=4.cm]{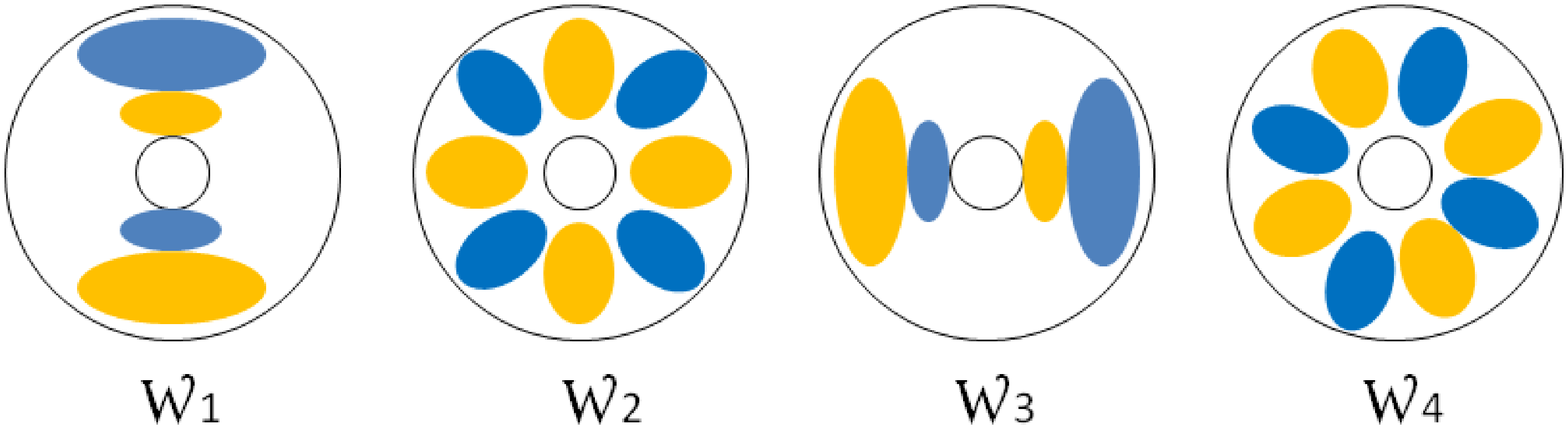}}
\vspace{.1cm}
\begin{caption}{\sl Schematic representation of four independent
eigenfunctions on the annular domain having the same eigenvalue.}
\end{caption}
\end{figure}
\end{center}

\vfill

\begin{center}
\vspace{-.1cm}
\begin{figure}[p]
\centerline{\includegraphics[width=5.cm,height=3.8cm]{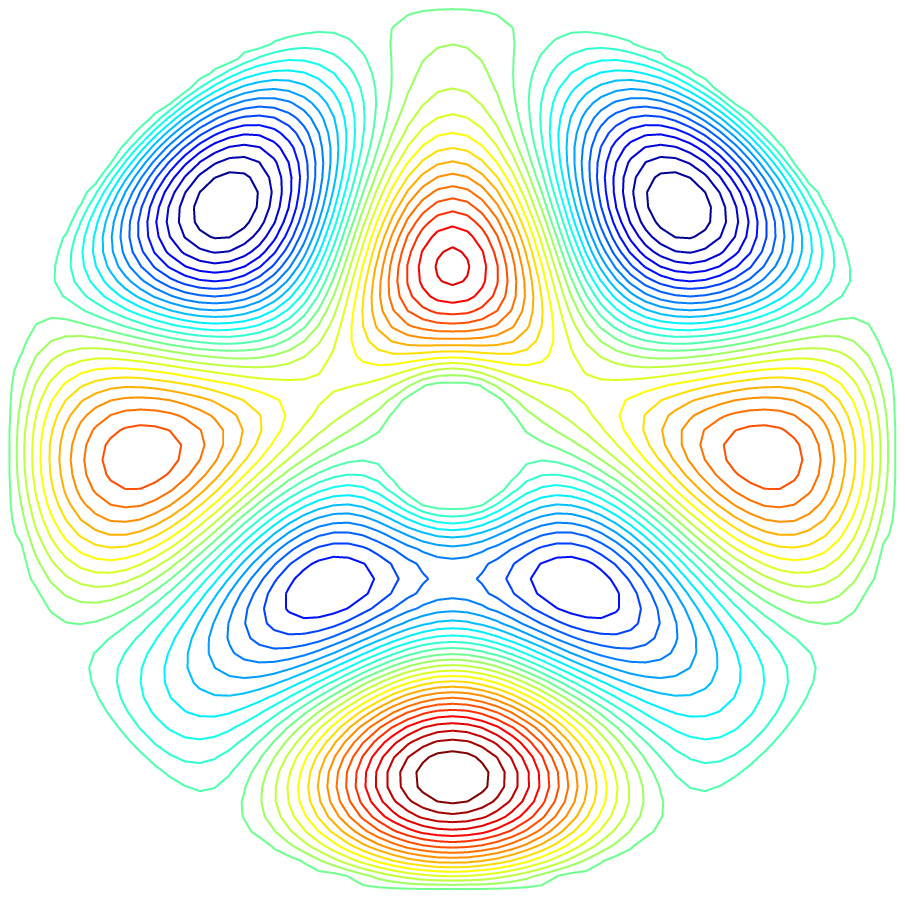}
\hspace{-1.cm}\includegraphics[width=5.cm,height=3.8cm]{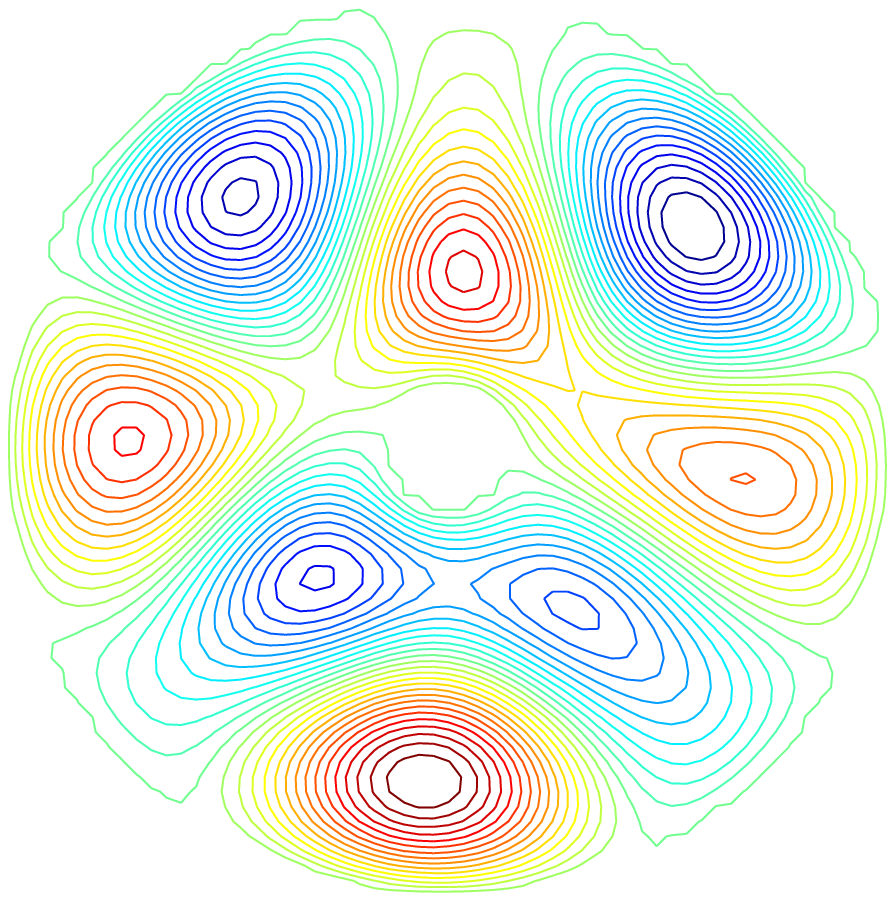}
\hspace{-1.cm}\includegraphics[width=5.cm,height=3.8cm]{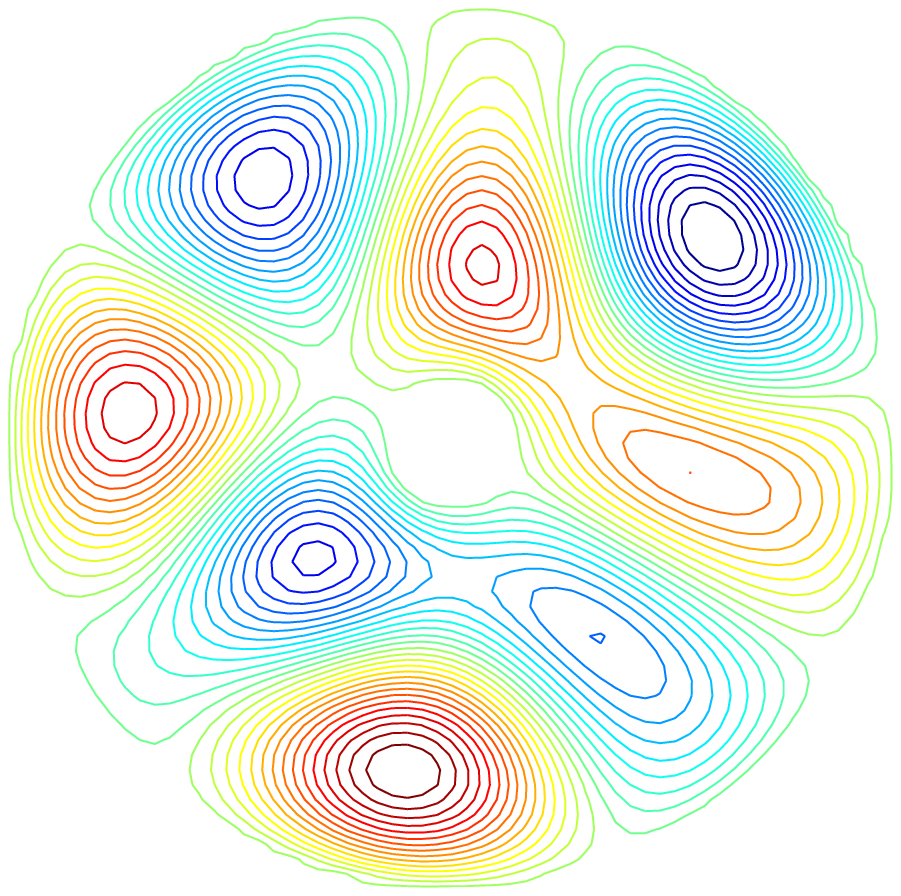}}
\vspace{.3cm}
\centerline{\includegraphics[width=5.cm,height=3.8cm]{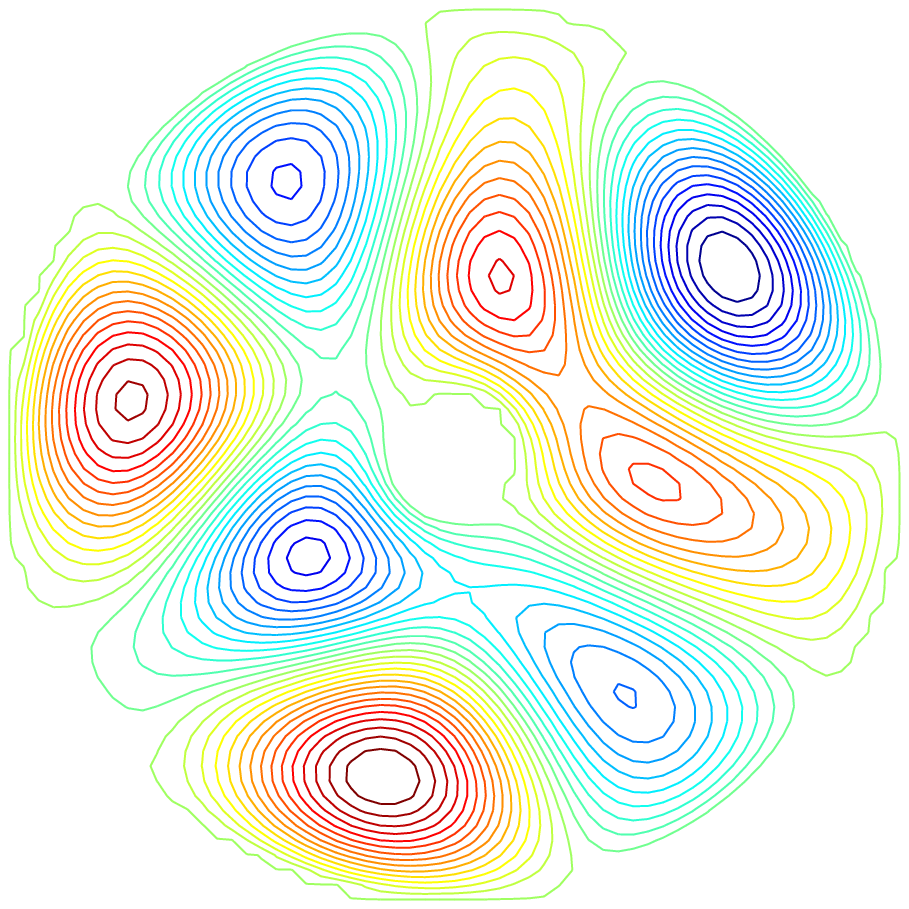}
\hspace{-1.cm}\includegraphics[width=5.cm,height=3.8cm]{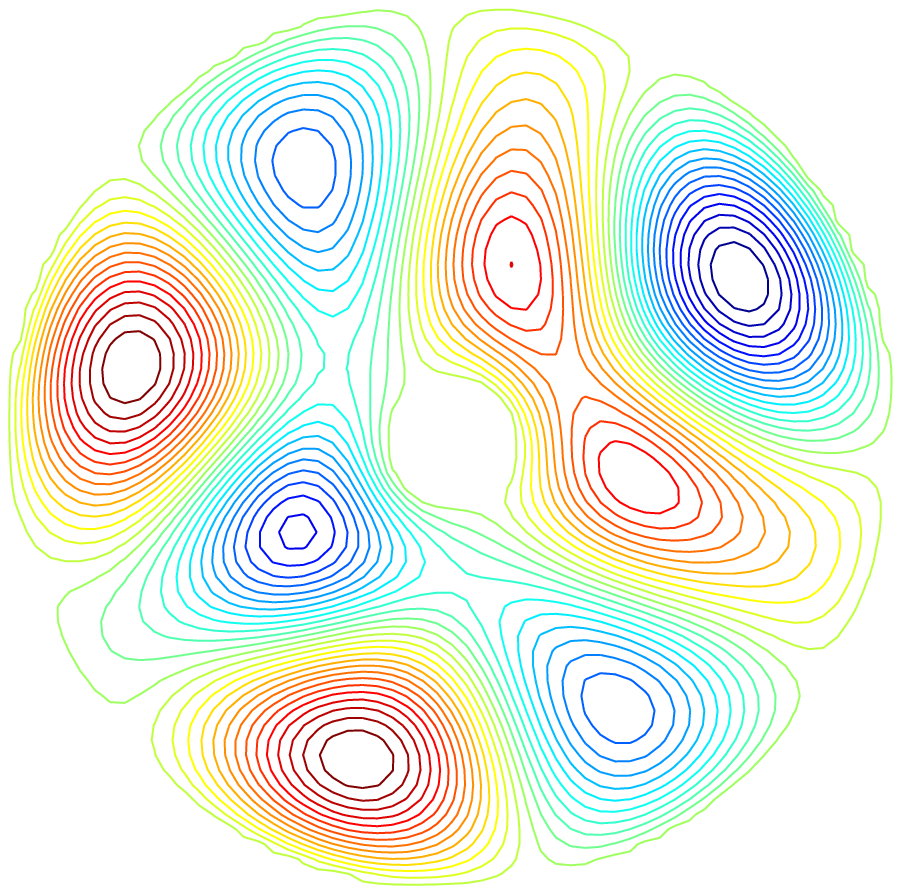}
\hspace{-1.cm}\includegraphics[width=5.cm,height=3.8cm]{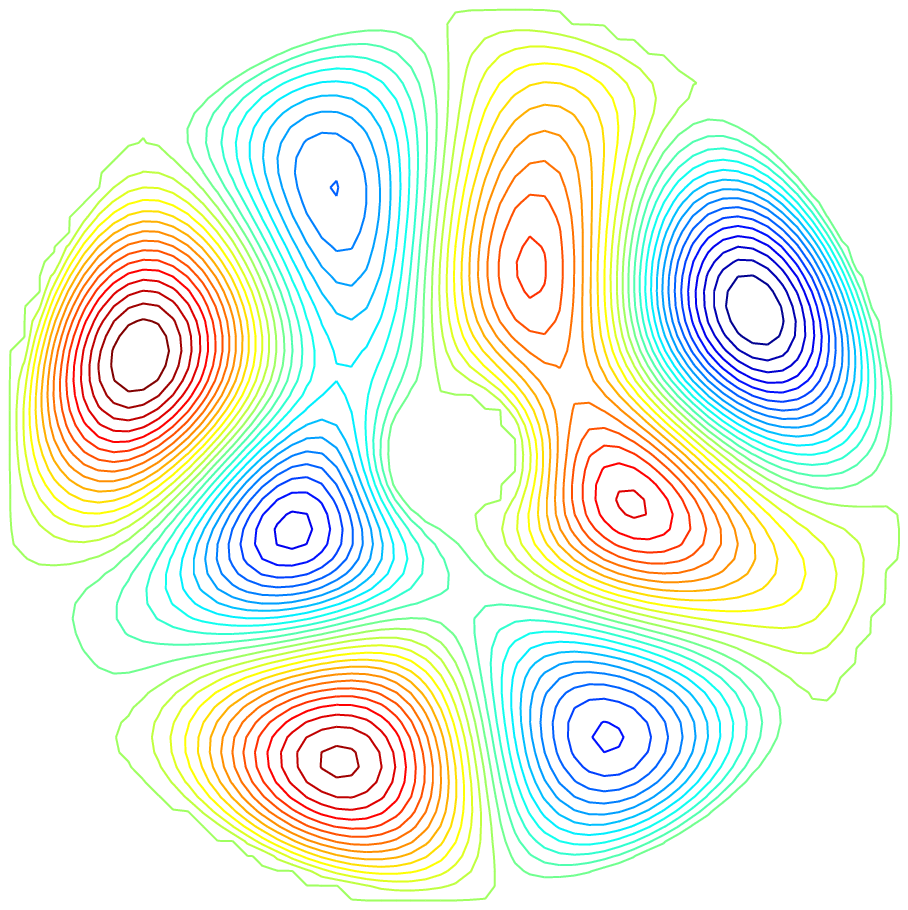}}
\vspace{.3cm}
\centerline{\includegraphics[width=5.cm,height=3.8cm]{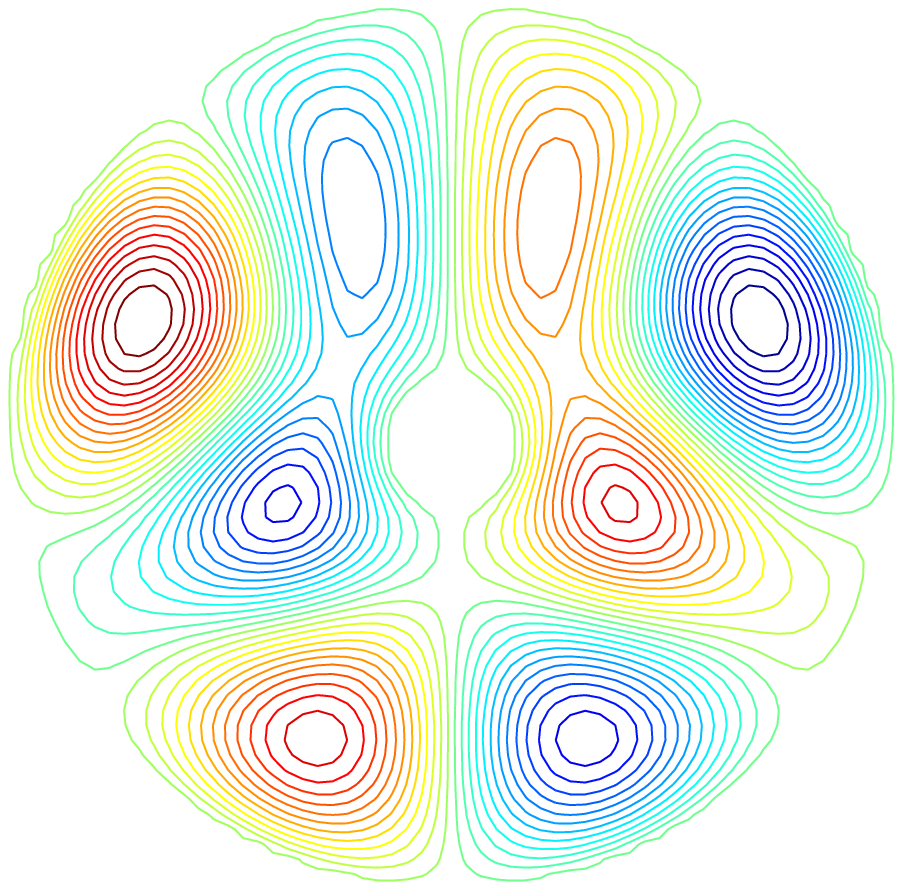}
\hspace{-1.cm}\includegraphics[width=5.cm,height=3.8cm]{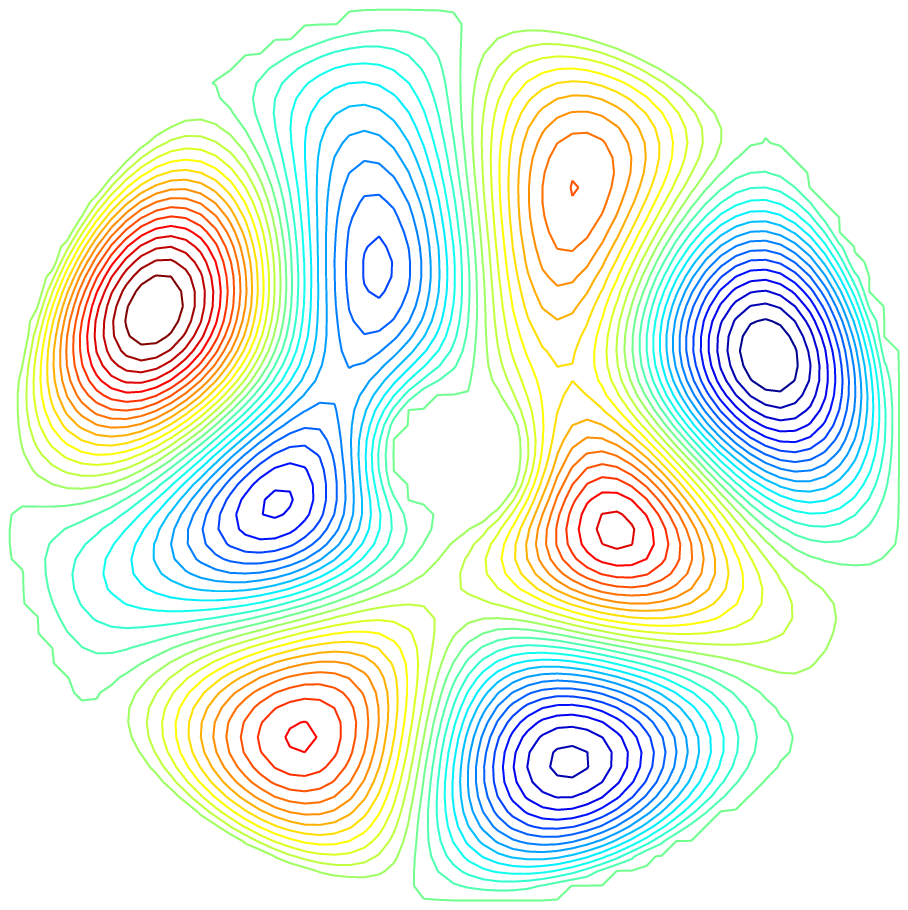}
\hspace{-1.cm}\includegraphics[width=5.cm,height=3.8cm]{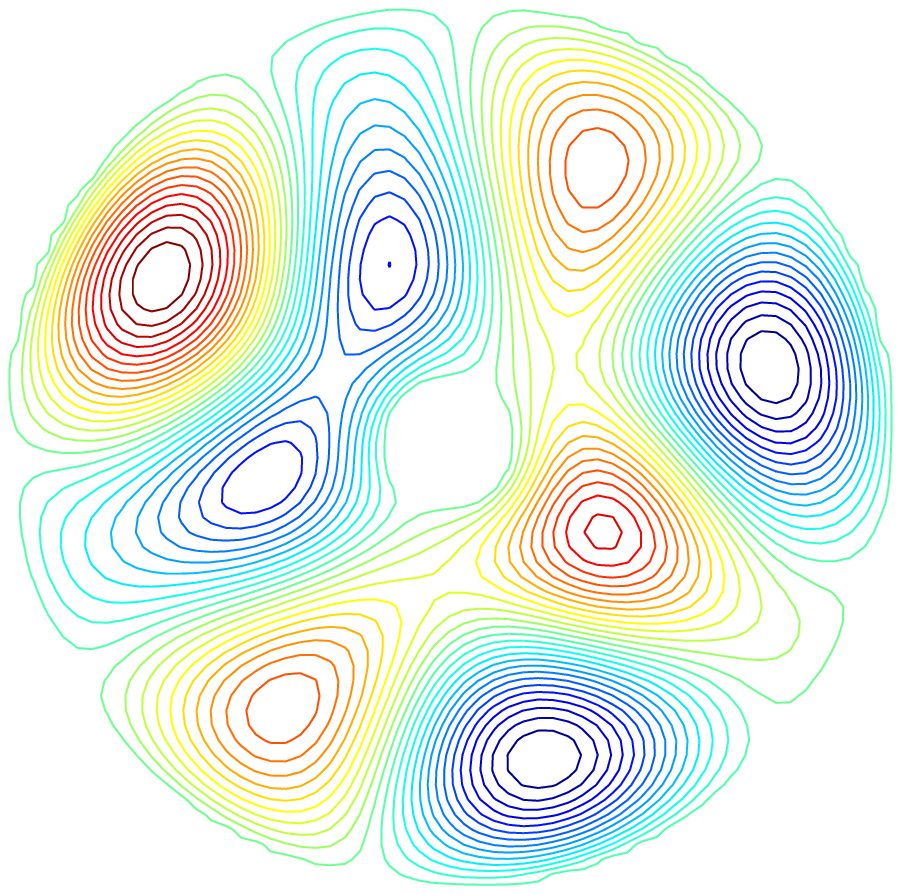}}
\vspace{.3cm}
\centerline{\includegraphics[width=5.cm,height=3.8cm]{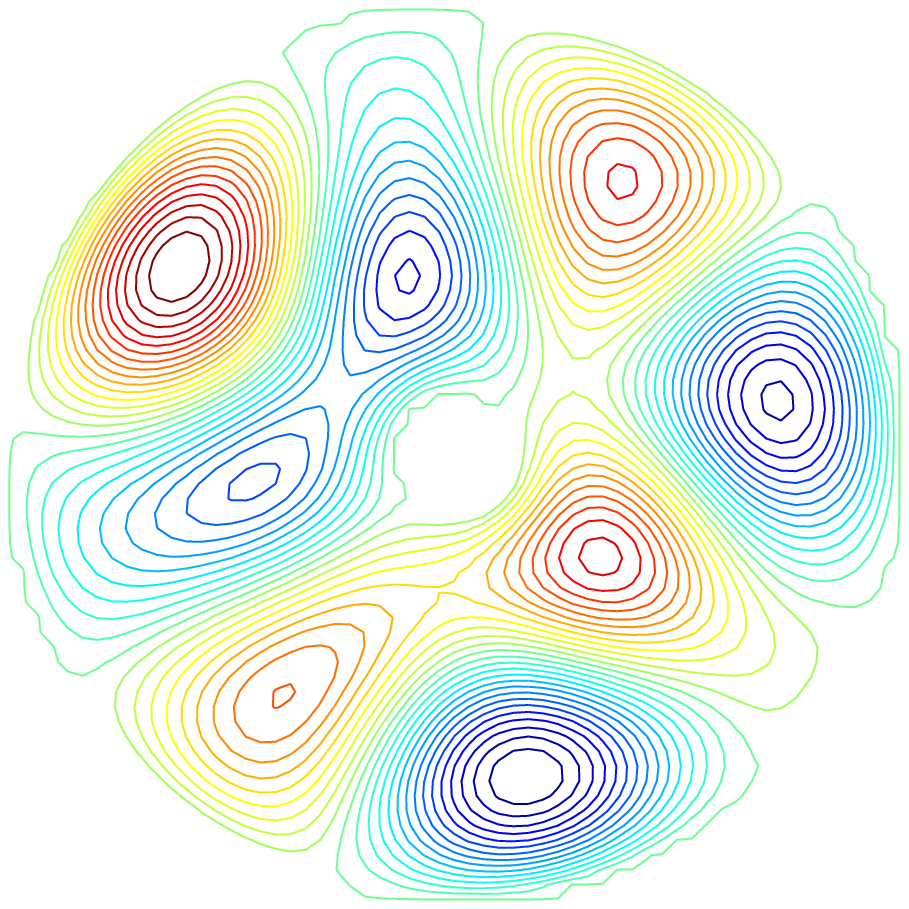}
\hspace{-1.cm}\includegraphics[width=5.cm,height=3.8cm]{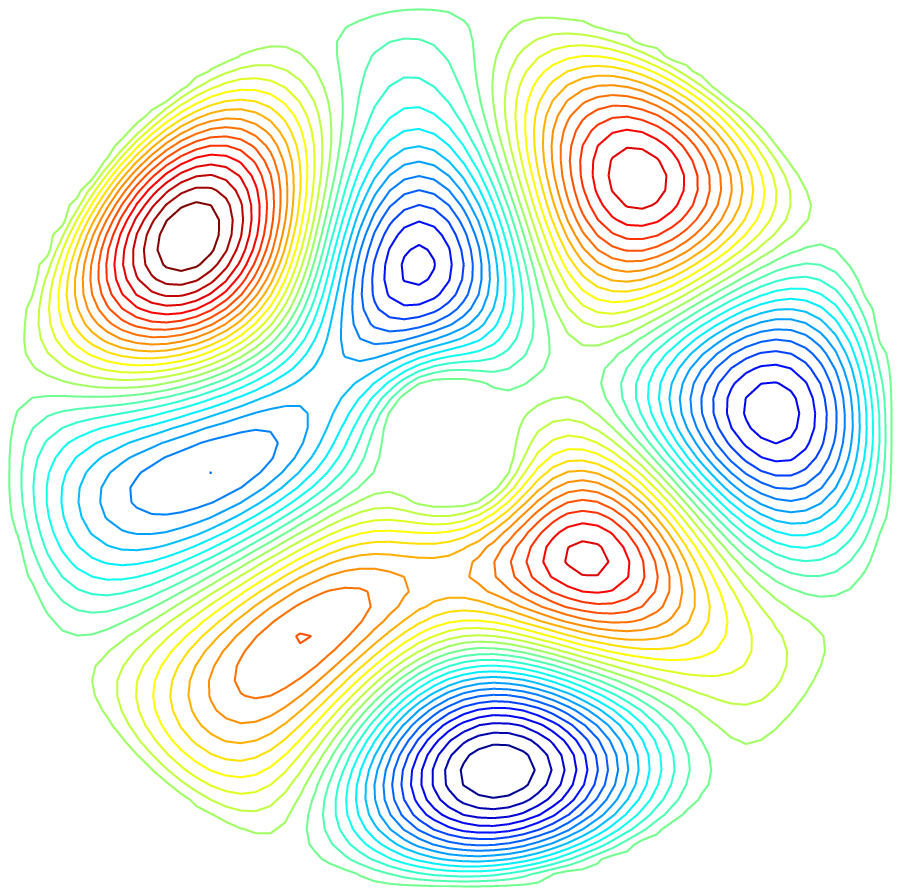}
\hspace{-1.cm}\includegraphics[width=5.cm,height=3.8cm]{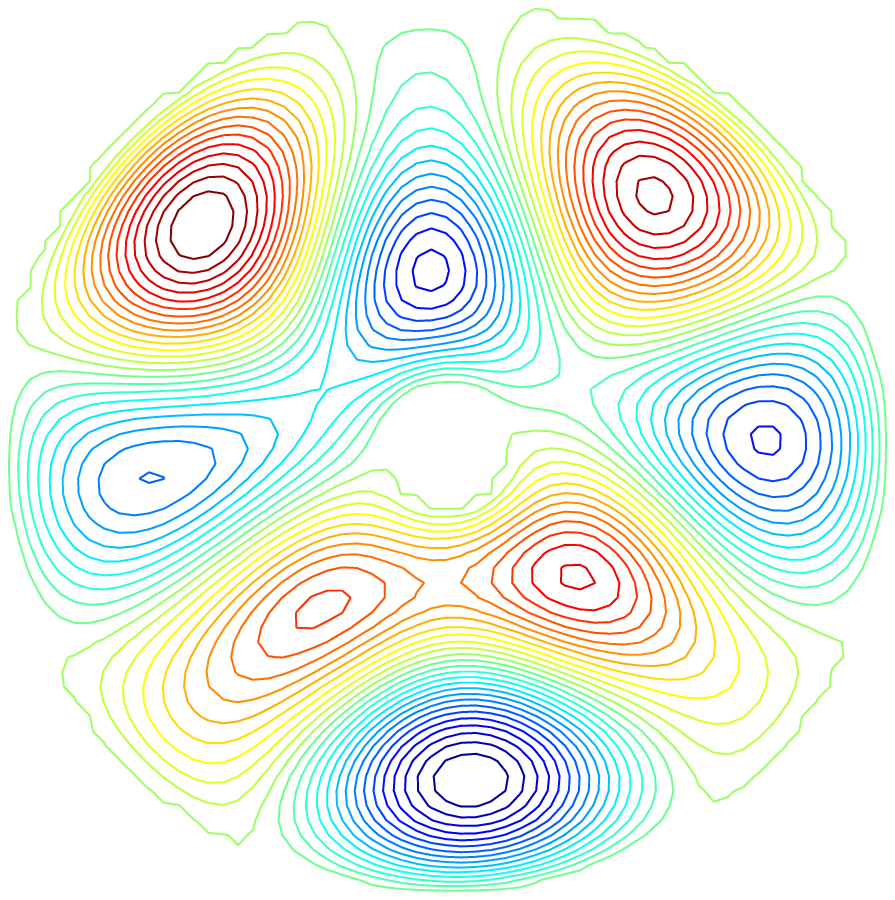}}
\vspace{.1cm}
\begin{caption}{\sl Evolution at different times of a linear combination
of eigenfunctions corresponding to the same eigenvalue. The two
eigenfunctions of Fig. 3 merge and rotate with a periodic behavior.
The frequency is uniquely established by the global size of the domain.}
\end{caption}
\end{figure}
\end{center}\vspace{.3cm}

\section{Conclusions}

From the theoretical viewpoint, Maxwell's equations allow
for special solutions trapped inside
an infinite  cylinder or, more or less equivalently, in a toroid.
The study here developed in vacuum reveals original configurations and may
help for instance in the analysis of confined plasma (see, e.g.,
\cite{hazeltine}). Other applications may be found in the field of the so called
whispering gallery resonators. Waves trapped in these cavities are
smoothly guided to circulate around by continuous reflection returning at
the origin with the initial phase. Spherical, cylindrical and ring-shaped
whispering galleries are commonly produced for a broad range
of industrial applications. Typical areas of interest are in fiber
telecommunications or biosensing.  The literature is very rich. We just
mention a couple of non extremely specialized publications: \cite{oraevsky},
\cite{snyder}.

\end{document}